\documentclass[twocolumn]{aastex631}

\usepackage{upgreek}
\usepackage{textcomp}
\usepackage{wasysym}
\usepackage{blindtext}
\usepackage{lineno}

\shorttitle{Wide-field Retrieval of Astrodata Program}
\shortauthors{Hunter Brooks}

\graphicspath{{./}{figures/}}

\begin{document}

\title{WRAP: A Tool for Efficient Cross-Identification of Proper Motion Objects Spanning Multiple Surveys }

\author[0000-0002-5253-0383]{Hunter Brooks}
\affiliation{Department of Astronomy and Planetary Science, Northern Arizona University, Flagstaff, AZ 86011, USA}
\affiliation{IPAC, Mail Code 100-22, Caltech, 1200 E. California Blvd., Pasadena, CA 91125, USA}

\author[0000-0003-4269-260X]{J.\ Davy Kirkpatrick}
\affiliation{IPAC, Mail Code 100-22, Caltech, 1200 E. California Blvd., Pasadena, CA 91125, USA}

\author[0000-0001-7896-5791]{Dan Caselden}
\affiliation{Department of Astrophysics, American Museum of Natural History, Central Park West at 79th Street, New York, NY 10024, USA}
\affiliation{Backyard Worlds: Planet 9}

\author[0000-0002-6294-5937]{Adam C. Schneider}
\affiliation{United States Naval Observatory, Flagstaff Station, 10391 West Naval Observatory Rd., Flagstaff, AZ 86005, USA} 

\author[0000-0002-1125-7384]{Aaron M. Meisner}
\affiliation{NSF's National Optical-Infrared Astronomy Research
Laboratory, 950 N. Cherry Ave., Tucson, AZ 85719, USA}

\author[0000-0001-9778-7054]{Yadukrishna Raghu}
\affiliation{Backyard Worlds: Planet 9}

\author[0009-0009-1388-1541]{Farid Cedeno}
\affiliation{Backyard Worlds: Planet 9}

\author[0000-0001-6251-0573]{Jacqueline K. Faherty}
\affiliation{Department of Astrophysics, American Museum of Natural History, Central Park West at 79th Street, New York, NY 10024, USA}

\author[0000-0001-7519-1700]{Federico Marocco}
\affiliation{IPAC, Mail Code 100-22, Caltech, 1200 E. California Blvd., Pasadena, CA 91125, USA}

\author[0000-0002-2387-5489]{Marc J. Kuchner}
\affiliation{NASA Goddard Space Flight Center, Exoplanets and Stellar Astrophysics Laboratory, Code 667, Greenbelt, MD 20771, USA}

\author[0000-0003-2478-0120]{S.L.Casewell}
\affiliation{School of Physics and Astronomy, University of Leicester, University Road, Leicester, LE1 7RH, UK}

\author{The Backyard Worlds:  Planet 9 Collaboration}

\begin{abstract}
We introduce the Wide-field Retrieval of Astrodata Program (WRAP), a tool created to aid astronomers in gathering photometric and astrometric data for point sources that may confuse simple cross-matching algorithms because of their faintness or motion. WRAP allows astronomers to correctly cross-identify objects with proper motion across multiple surveys by wedding the catalog data with its underlying images, thus providing visual confirmation of cross-associations in real time. Developed within the Backyard Worlds: Planet 9 citizen science project, WRAP aims to aid in the characterization of faint, high motion sources by this collaboration (and others).
\end{abstract}


\section{Introduction}\label{intro}
Gathering catalog photometry and astrometry is generally straight-forward, although the process becomes notably challenging when dealing with faint or high proper motion objects. There are multiple ways to gather this information, like Aladin (\citealt{Bonnarel et al.(2000)}), CDS Cross-Match Service (\citealt{Boch et al.(2012)}), etc. We built upon previous techniques to develop the Wide-field Retrieval of Astrodata Program (WRAP) to improve the process. Other techniques have several drawbacks, such as complex GUI interfaces, a need for understanding multiple programming languages, or a lack of visual verification. WRAP is designed to address these challenges. We start by examining previous techniques and programs in \S\ref{prev tech}. Following that, we leverage the advantages of these prior methods to develop WRAP in \S\ref{WRAP}.

\section{Previous Techniques} \label{prev tech}

\begin{figure*}[ht]
        \centering
        \includegraphics[scale = 0.35]{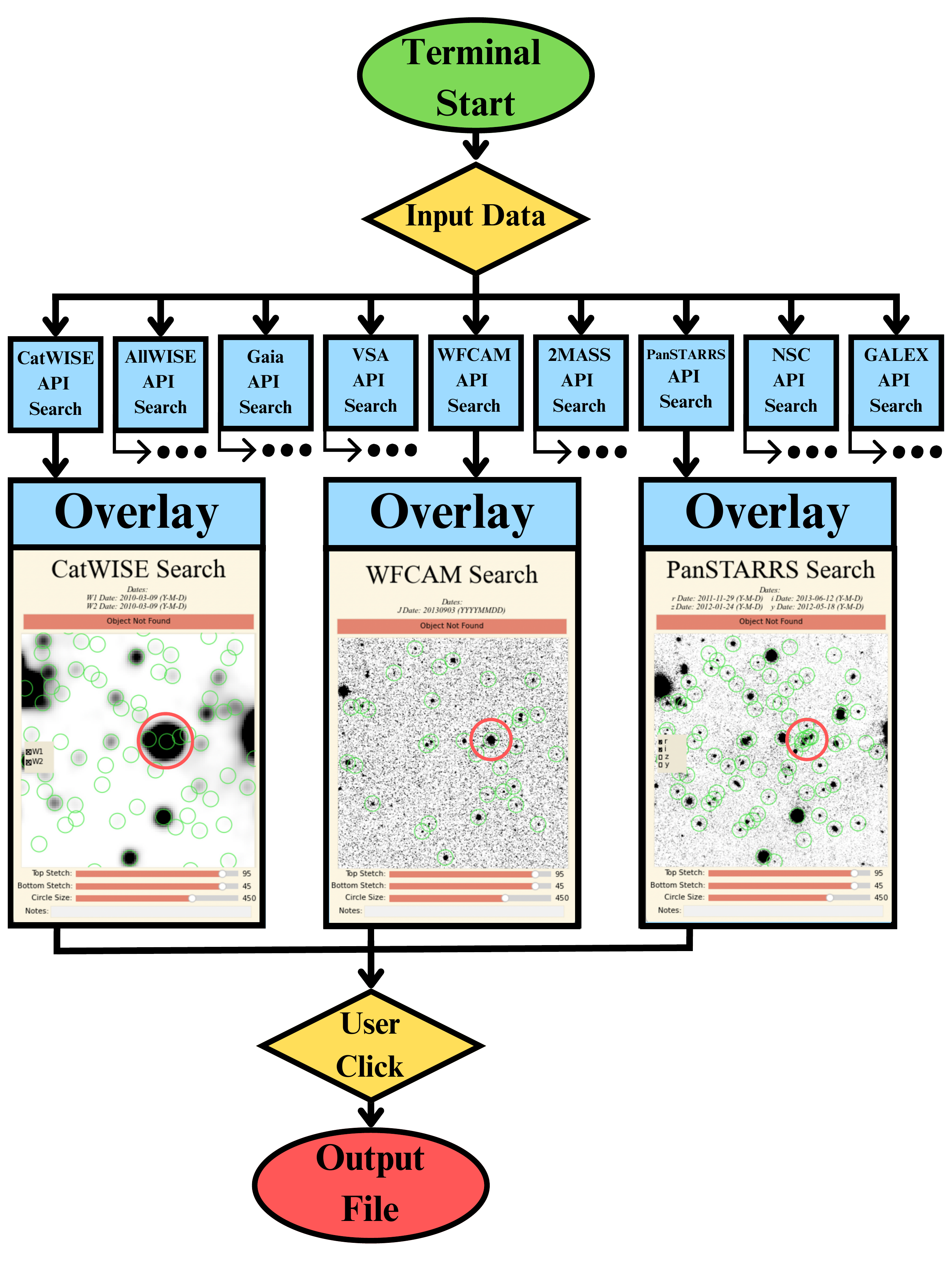}
        \caption{A flowchart illustrating the WRAP procedure. Here we utilize 2MASSI J1721039+334415 (\citealt{Bardalez Gagliuffi et al.(2014)}). It is highlighted in red for CatWISE, WFCAM, and PanSTARRS. The process begins by opening WRAP in a terminal window. Then, the user inputs the coordinates, sets the search radius, and chooses the catalogs for their search. WRAP executes API queries, interacting with the chosen catalog APIs to retrieve relevant data. If the designated region is found through the API, WRAP displays an image of the area with catalog detections overlaid. Users interact by clicking on the image, prompting WRAP to identify the nearest object and writing pertinent information from the catalog into an output file. }
        \label{Figure 1}
\end{figure*}

Numerous additional tools have been developed to facilitate the gathering of astrophysical data. However, discussing every software in detail here would be too enormous a task for a research note. Instead, we have highlighted some of the most popular techniques and software currently available to provide context for understanding the improvements WRAP provides.

The most prevalent approach for collecting photometric and astrometric data involves conducting cross-matches based on Right Ascension (R.A.) and Declination (Decl.). One such program is the CDS X-Match Service (\citealt{Boch et al.(2012)}; \citealt{Pineau et al.(2020)}) allowing users to search various catalogs using an input coordinate table.  However, a drawback is the absence of visual confirmation for matched data. This makes it susceptible to errors from incorrect cross-matches if the object in question has large proper motion. In contrast, other services offer visual confirmation, enhancing the accuracy of gathering astrophysical data.

The Aladin catalog search feature (\citealt{Bonnarel et al.(2000)}) shares similarities with WRAP in terms of functionality. It enables users to load images and superimpose them onto the corresponding source catalog. However, this method can be hampered by slow performance, primarily when dealing with substantial amounts of data simultaneously. Moreover, this approach is primarily designed for single object searches, limiting its applicability for more extensive cross-identification tasks.

Additional software includes the IRSA Viewer (\url{https://irsa.ipac.caltech.edu/irsaviewer}), VizieR Photometry Viewer (\citealt{Ochsenbein et al.(2000)}), and AstroToolBox (also developed by the Backyward Worlds: Planet 9 collaboration; \citealt{Kiwy(2022)}). However, these tools have certain drawbacks,  briefly described here. The IRSA Viewer, for instance, lacks catalogs beyond those supported at IPAC, limiting the available data. The VizieR Photometry Viewer is comparable to the CDS X-Match Service in that it lacks readily available imaging for visual confirmation. Lastly, AstroToolBox offers both visual confirmation and cross-matching, but these features are located in separate tabs.

\section{WRAP Overview} \label{WRAP}
To address these complications, we developed WRAP. Developed in Python 3.8 (the recommended version for using WRAP), WRAP is an open-source application designed to have minimal dependencies and low computational expenses, ensuring it remains freely accessible. After installation, the user inputs single or multi-object files (in CSV, FITS, ASCII, or IPAC table formats): R.A. (deg), Decl. (deg), search radius (arcsec), then selects the desired catalogs. The catalogs included are: CatWISE2020 Point Source Catalog (\citealt{Marocco et al.(2021)}), AllWISE Source Catalog (WISE; \citealt{Cutri et al.(2021)}), Gaia Data Release 3 (Gaia; \citealt{Gaia Collaboration et al.(2021)}), VISTA Science Archive catalogs (VSA; \citealt{Irwin et al.(2004)}), WFCAM Science Archive catalogs (WFCAM; \citealt{Hambly et al.(2008)}), 2MASS Point Source Catalog (2MASS; \citealt{Cutri et al.(2003)} and \citealt{Skrutskie et al.(2006)}), PanSTARRS Data Release 2 (PanSTARRS; \citealt{Chambers et al.(2016)}), NOIRLab Source Catalog Data Release 2 (NSC; \citealt{Nidever et al.(2021)}), and the GALEX Catalog (GALEX; \citealt{Martin et al.(2005)}). 

Once inputs are selected, a WISEView (\citealt{Caselden et al. 2018}) window will display WISE coadd images to aid the user in finding their object. WRAP begins by running through each catalog, showing the images with overlaid sources. Each catalog overlay comes with customizable image features, such as image stretch, photometric band selection, and circle size. Furthermore, users may add notes to any selected catalog.  After the user clicks on their chosen object, WRAP identifies the nearest object through a proximity check within the field, and the results are then saved to a .csv table (no images are saved in WRAP).

In Figure \ref{Figure 1}, the flowchart demonstrates the intuitive functionality of WRAP, emphasizing its user-friendly interface. The catalog overlay feature of WRAP is essential for confirmation of high motion objects, such as the illustrated object with its high proper motion of $\sim$1.95 arcsec yr$^{-1}$ (\citealt{Bardalez Gagliuffi et al.(2014)}), that would be difficult to cross-identify through a simple positional cross-match. 

\section{Conclusion} \label{conclusion}
WRAP was developed to serve as a tool for astronomers seeking photometric and astrometric data for faint and high proper motion objects. It incorporates the most effective features from previous methods, resulting in a user-friendly and streamlined process. This is achieved by offering visual verification, consolidating popular catalogs in one location, and requiring no coding experience. WRAP is accessible from GitHub at the following link: \url{https://github.com/huntbrooks85/WRAP}. The DOI for WRAP is: \url{https://doi.org/10.5281/zenodo.10359982}. Please refer to the ReadMe file on the GitHub page for detailed installation and usage instructions. 

\section{Acknowledgements}

This Backyard Worlds research was supported by NASA grant 2017-ADAP17-0067. All PanSTARRS data used in this paper can be found in MAST: \dataset[10.17909/s0zg-jx37]{http://dx.doi.org/10.17909/s0zg-jx37}

\software{WiseView (\citealt{Caselden et al. 2018})}
\facilities{NEOWISE, Gaia, ESO:VISTA, UKIRT, CTIO:2MASS, FLWO:2MASS, PS2, Astro Data Lab, IRSA, CDS, MAST}

\end{document}